\documentclass[a4paper,12pt]{extarticle}
\pdfoutput=1

\usepackage[english]{babel}
\usepackage[utf8x]{inputenc}
\usepackage[T1]{fontenc}

\usepackage[a4paper,top=3cm,bottom=2cm,left=2cm,right=2cm,marginparwidth=1.75cm]{geometry}

\usepackage{natbib}
\usepackage{subfig}
\usepackage{amsmath}
\usepackage{graphicx}
\usepackage{bm}
\usepackage{lipsum}
\usepackage{mwe}
\usepackage[colorinlistoftodos]{todonotes}
\usepackage[colorlinks=true, allcolors=blue]{hyperref}

\title{Power Comparisons in 2x2 Contingency Tables: Odds Ratio versus Pearson Correlation versus Canonical Correlation}

\author{Mohammad Alfrad Nobel Bhuiyan, Michael Wathen, Marepalli Rao }

\begin{document}

\maketitle

\begin{abstract}
It is an important inferential problem to test no association between two binary variables based on data. Tests based on the sample odds ratio are commonly used. We bring in a competing test based on the Pearson correlation coefficient. In particular, the Odds ratio does not extend to higher order contingency tables, whereas Pearson correlation does. It is important to understand how Pearson correlation stacks against the odds ratio in 2x2 tables. Another measure of association is the canonical correlation. In this paper, we examine how competitive Pearson correlation is vis a vis odds ratio in terms of power in the binary context, contrasting further with both the Wald Z and Rao Score tests. We generated an extensive collection of joint distributions of the binary variables and estimated the power of the tests under each joint alternative distribution based on random samples. The consensus is that none of the tests dominates the other.

keywords: Odds ratio, Pearson correlation, Canonical correlation 
\end{abstract}

\section{Introduction}
Let X  and Y  be two  binary random variables with joint distribution,
$$
Q=
\begin{pmatrix}
p_{11} & p_{12}  \\
p_{21} & p_{22} \\
\end{pmatrix}
$$
\newline
Let the marginal probabilities be $p_{1+}$, $p_{2+}$, $p_{+1}$, $p_{+2}$.
The odds ratio is defined by,
$$ \theta =  \frac{p_{11}  p_{22}} {p_{12}  p_{21}} $$\\
which is a measure of association between X and Y.\\
Assumptions: $\colon$
\begin{itemize}
    \item  $0\leq \theta \leq \infty$
    \item  X and Y are independent if and only if $\theta$ = 1
    \item  Odds ratio measures to what extent the variables are away from independence.
    \item  The ratio $\theta \geq 1$ means 
$Pr(X = 1, Y = 1) > Pr(X = 1) Pr(Y = 1)$. It is more likely to get $X = 1$ and $Y = 1$ than is possible under independence. 
\end{itemize}
The joint distribution is unknown. Our test of Hypothesis is,
Null Hypothesis $(H_{0})$: X and Y are independent. \\
\hspace*{3 cm} vs \\ 
Alternative Hypothesis $(H_{1})$: X and Y are not independent.\\
\newline
Both null and alternative hypotheses are composite. Several tests can be built based on a random sample
$
\begin{pmatrix}
n_{11} & n_{12}  \\
n_{21} & n_{22} \\
\end{pmatrix}
$
from the joint distribution.

\subsection{Tests based on sample odds ratio}
The likelihood estimator of $\theta$ is given by
$$ \widehat{\theta} =  \frac{n_{11}  n_{22}} {n_{12}  n_{21}} $$
Let the marginal totals be  $n_{1+}$, $n_{2+}$, $n_{+1}$, and $n_{+2}$.  The asymptotic variance of $ln \left( \widehat{\theta} \right )$ is given (Courtesy: Delta method \citep{cox2005delta}, \citep{agresti2003categorical}, \citep{agresti2010analysis}) by 

\[{Asy \rm{Var}}\left( {\ln \widehat \theta } \right) = \frac{1}{{n{p_{11}}}} + \frac{1}{{n{p_{12}}}} + \frac{1}{{n{p_{21}}}} + \frac{1}{{n{p_{22}}}}\]
and it is estimated by
\[\widehat {{\mathop{Asy\rm Var}\nolimits} }\left( {\ln \widehat \theta } \right) = \frac{1}{{{n_{11}}}} + \frac{1}{{{n_{12}}}} + \frac{1}{{{n_{21}}}} + \frac{1}{{{n_{22}}}}\]
The Wald’s Z-statistic is given by
\begin{equation}
{Z_1} = \frac{{\ln \widehat \theta  - \ln \theta }}{{\sqrt {\widehat {AsyVar}\left( {\ln \widehat \theta } \right)} }}   
\end{equation}
\newline
which has a standard normal distribution, in large samples.  In particular, 
${Z_1} = \frac{{\ln \hat \theta  }}{{\sqrt {Var\left( {\ln \hat \theta } \right)} }}$
has the standard normal distribution N(0, 1) under the null hypothesis. 
An alternative to the Wald statistic is Rao’s Score statistic. The variance of $\ln \widehat \theta $  is calculated under the null hypothesis and then estimated. The statistic is given by
\begin{equation}
 {Z_2} = \frac{{\ln \hat \theta }}{{\sqrt {{{\widehat {Asy Var}}_{{H_o}}}\left( {\ln \hat \theta } \right)} }}   
\end{equation}
\newline
The statistic $Z_2$ has a standard normal distribution under the null hypothesis for large samples. The formula for the asymptotic variance is given by:
$$Va{r_{{H_o}}}\left( {\ln \hat \theta } \right) = \frac{1}{{n{p_{1 + }}{p_{ + 1}}}} + \frac{1}{{n{p_{1 + }}{p_{ + 2}}}} + \frac{1}{{n{p_{2 + }}{p_{ + 1}}}} + \frac{1}{{n{p_{2 + }}{p_{ + 2}}}}$$
and it is estimated by
$${\widehat {Asy Var}_{{H_o}}}\left( {\ln \hat \theta } \right) = \frac{n}{{{n_{1 + }}{n_{ + 1}}}} + \frac{n}{{{n_{1 + }}{n_{ + 2}}}} + \frac{n}{{{n_{2 + }}{n_{ + 1}}}} + \frac{n}{{{n_{2 + }}{n_{ + 2}}}}.$$
Of course, we could have used the traditional chi-squared statistic for testing independence. However, unlike the odds ratio, there is no population chi-squared measure of association. We will relate  the chi-squared statistic to the likelihood estimate of Pearson correlation in Section 1.2. 
\subsection{Tests based Pearson correlation}
We are looking for a competitor to the odds ratio. One competitor is the Pearson Correlation \citep{hayes1963statistics}. The population correlation is given by 
\[\phi  = \frac{{{p_{11}}{p_{22}} - {p_{12}}{p_{21}}}}{{\sqrt {{p_{1 + }}{p_{2 + }}{p_{ + 1}}{p_{ + 2}}} }},\]
Where the entities under the square root are the marginal probabilities and it has the property $ - 1 \le \phi  \le 1$. The random variables $X$ and $Y$ are independent if and only if $\phi  = 0$.
The likelihood estimate of $\phi$ is given by
\[\widehat \phi  = \frac{{{n_{11}}{n_{22}} - {n_{12}}{n_{21}}}}{{\sqrt {{n_{1 + }}{n_{2 + }}{n_{ + 1}}{n_{ + 2}}} }}\]
A Z-statistic a la Wald can be built based on the likelihood estimator $\widehat \phi $ of $\phi $. For the record, it is spelled out by
\begin{equation}
{Z_3} = \frac{{\widehat \phi }}{{\sqrt {\widehat {AsyVar}\left( {\widehat \phi } \right)} }}. 
\end{equation}
\newline
The asymptotic variance of $\widehat\phi$ a la the delta method is given in Appendix 1. For a description of the delta method, see \citep{cox2005delta}.
Another competitor is the canonical correlation \citep{lancaster20020} defined by
\[\rho  = \sqrt {{p_{1 + }}{p_{ + 1}}{p_{2 + }}{p_{ + 2}}} \left[ {\frac{{{p_{11}}}}{{{p_{1 + }}{p_{ + 1}}}} - \frac{{{p_{12}}}}{{{p_{1 + }}{p_{ + 2}}}} - \frac{{{p_{21}}}}{{{p_{2 + }}{p_{ + 1}}}} + \frac{{{p_{22}}}}{{{p_{2 + }}{p_{ + 2}}}}} \right]\]
It can be checked that $\phi = \rho$. We use the notation $\phi$ and  $\rho$ interchangeably.Our motivation for roping in the canonical correlation into the mix goes a bit deeper. Canonical correlations arise from the singular value decomposition of a transform of the joint distribution. Several layers of dependence between $X$ and $Y$ shine through (singular values) the canonical correlations. In the $2 x 2$, there is only one canonical correlation $\rho$ and it is exactly the same as the Pearson $\phi$. 
\newline
As an alternative to Wald’s $Z$ statistic, we have Rao’s score statistic based on $\widehat\phi$
\begin{equation}
{Z_4} = \frac{{\widehat \phi }}{{\sqrt {{{\widehat {AsyVar}}_{{H_o}}}\left( {\widehat \phi } \right)} }}.  
\end{equation}
\newline
It turns out that, $AsyVa{r_{{H_o}}}\left( {\widehat \phi } \right) = \frac{1}{n}$ \citep{o1981note}.
It can be checked that $n\widehat \phi^2 = {\chi ^2}$  (\citep{o1978asymptotic},\citep{o1978distributional}), the usual chi-squared statistic of the data in the $2 x 2$ contingency table \citep{hayes1963statistics}. We set the level of significance at $5\% $. Reject the null hypothesis at $5\% $ level of significance if $\left| {{Z_i}} \right| > 1.96$.
\begin{flushleft}
The goals in this work are now spelled out.
\end{flushleft}
\begin{itemize}
	\item Compare and contrast the properties of the measures of association: $\phi$ and $\theta$ (Sections 2 and 3).
	\item Make power comparisons between the Wald’s test ($Z_1$) and Rao’s score test ($Z_2$) based on the odds ratio, Wald’s test ($Z_3$) and Rao’s score test ($Z_4$), which is the same as the chi-squared test, based on the Pearson correlation or canonical correlation (Section 4).
\end{itemize}
Power comparisons made via extensive simulations.
\begin{enumerate}
    \item Draw randomly 100 distributions from the space $\Omega  = \left\{ {\left( {{p_{11,}}{p_{12,}}{p_{21,}}{p_{22}}} \right);{p_{ij}} \ge 0,sum = 1} \right\}.$ For sampling, we use the uniform  Dirichlet distribution: ${\mathop{\rm Dirichlet}\nolimits} \left( {{p_{11}},{p_{12}},{p_{21}},{p_{22}};1,1,1,1} \right)$ whose joint density is given by $f\left( {{p_{11}},{\rm{ }}{p_{12}},{\rm{ }}{p_{21}},{\rm{ }}{p_{22}}} \right){\rm{ }} = {\rm{ }}6$,  $\left( {{p_{11}},{\rm{ }}{p_{12}},{\rm{ }}{p_{21}},{\rm{ }}{p_{22}}} \right) \in \Omega$. Marginally,
    ${p_{ij}}$s are identically distributed. The marginal distribution of $p_{11}$ is ${\rm Beta}\left( {1,3} \right)$ with
    ${\rm E}\left( {{p_{11}}} \right) = {\textstyle{1 \over 4}}$ and
    ${\mathop{\rm Var}\nolimits} \left( {{p_{11}}} \right) = {\textstyle{3 \over {80}}}$.
    \item With probability one, under each joint distribution, $X$ and $Y$ are associated.
    \item From each joint distribution $({p_{11}},{\rm{ }}{p_{12}},{\rm{ }}{p_{21}},{\rm{ }}{p_{22}})$ generated from the uniform Dirichlet distribution, generate a random sample 
    $({n_{11}},{\rm{ }}{n_{12}},{\rm{ }}{n_{21}},{\rm{ }}{n_{22}})$ of 100 observations from the ${\mathop{\rm Multinomial }\nolimits} ({n_{11}},{\rm{ }}{n_{12}},{\rm{ }}{n_{21}},{\rm{ }}{n_{22}};{\rm{ }}{\mathop{\rm prob}\nolimits} {\rm{ }} = {\rm{ }}\left( {{p_{11}},{\rm{ }}{p_{12}},{\rm{ }}{p_{21}},{\rm{ }}{p_{22}}}) \right)$. The reason we have chosen the sample size to be 100 is that we can reasonably expect each ${n_{ij}} \ge 5$. All the tests we are entertaining are asymptotic in nature and we are following the dictum stipulated by \citep{cochran1952chi2}, \citep{cochran1954some} for the applicability of the asymptotic tests. For each Multinomial sample, we apply all the four tests defined by  (1), (2), (3), and (4) at $5\%$ level of significance. We set up a counter for each test by: ${\mathop{\rm Counter}\nolimits}  = 1$ if the null hypothesis is rejected, $0$, if not rejected. Repeat the Multinomial sampling 1000 times. The estimated power under a test is the proportion of times the null hypothesis is rejected.
    \item Present the results by tables and graphs.
\end{enumerate}
In Section 2, we contrast Pearson $\phi$ and $ln (Odds ratio)$. In Section 3, we explain  the background of canonical correlation. In Section 4, we present the results. In Section 5, we discuss the results. The asymptotic variance of $\widehat{\phi}$ is presented in the Appendix 1.  

\section{Canonical Correlation (Pearson correlation) versus Odds Ratio}
A number of self-evident  truths are as follows, .
\begin{itemize}
    \item The event $\left\{ {X{\rm{ }} = {\rm{ }}1,{\rm{ }}Y{\rm{ }} = {\rm{ }}1} \right\}$ is more likely than under the independence of X and Y if and only if $\theta {\rm{ }} > {\rm{ }}1$ if and only if $ln \left( \theta {\rm{ }} \right ) > {\rm{ }}0$ if and only if $\rho {\rm{ }} > {\rm{ }}0$.
    \item The event $\left\{ {X{\rm{ }} = {\rm{ }}1,{\rm{ }}Y{\rm{ }} = {\rm{ }}1} \right\}$ is less likely than under the independence of X and Y if and only if $\theta {\rm{ }} < {\rm{ }}1$ if and only if $\ln \left( \theta  \right) < 0$ and if and only if $\rho {\rm{ }} < {\rm{ }}0$.
    \item $ - \infty  \le \ln \left (\theta \right ) \le \infty $
    \item $ - 1 \le \rho  \le 1$
    \item The correlations are more attractive in that their ranges are bounded. However, the odds ratio has better interpretability than the correlation.
    \item If the joint distribution is $\left( {\begin{array}{*{20}{c}}
{0.5}&0\\
0&{0.5}
\end{array}} \right)$, $\ln \left( \theta \right) = \infty $ and $\rho =1$.
    \item If the joint distribution is $\left( {\begin{array}{*{20}{c}}
0&{0.5}\\
{0.5}&0
\end{array}} \right)$, $\ln \left (\theta \right)  = -\infty $ and $\rho =-1$.
\end{itemize}
We introduce four pillars of the joint distribution: 
$A = \frac{{{p_{11}}}}{{{p_{1 + }}{p_{ + 1}}}}$ ; 
$B = \frac{{{p_{12}}}}{{{p_{1 + }}{p_{ + 2}}}}$ ;
$C = \frac{{{p_{21}}}}{{{p_{2 + }}{p_{ + 1}}}}$ ;
$D = \frac{{{p_{22}}}}{{{p_{2 + }}{p_{ + 2}}}}$.
Another characterization in terms of the pillars emerges as follows:
\[A > 1 \Leftrightarrow D > 1 \Leftrightarrow \theta  > 1 \Leftrightarrow \rho  > 0.\]
Let, ${G_1} = {\rm{ }}$ geometric mean of $A$ and $D = {\rm{ }}{\left( {AD} \right)^{0.5}}$, \\
\indent ${G_2} = {\rm{ }}$ geometric mean of $B$ and $C{\rm{ }} = {\rm{ }}{\left( {B*C} \right)^{0.5}}$, \\
\indent $A_{1}$ = Arithmetic mean of A and D = $\frac{A+D}{2}$, \\  
\indent $A_{2}$ = Arithmetic mean of B and C = $\frac{B+C}{2}$.  
\newline
The measures $\theta$ and $\rho$ are functions of these pillars through their arithmetic and geometric means. 
\newline
Odds ratio = $\theta= \left(\frac{G_{1}}{G_{2}}\right)^2$   and 
$ln\theta = 2(lnG_{1}-lnG_{2})$
\newline
The canonical correlation $\rho$ is connected to the pillars. 
\begin{equation*} 
\begin{split}
\rho  & = \sqrt {{p_{1 + }}{p_{ + 1}}{p_{2 + }}{p_{ + 2}}} \left[ {\frac{{{p_{11}}}}{{\sqrt {{p_{1 + }}{p_{ + 1}}} }} - \frac{{{p_{12}}}}{{\sqrt {{p_{1 + }}{p_{ + 2}}} }} - \frac{{{p_{21}}}}{{\sqrt {{p_{2 + }}{p_{ + 1}}} }} + \frac{{{p_{22}}}}{{\sqrt {{p_{2 + }}{p_{ + 2}}} }}} \right]\\
& = 2 \sqrt {{p_{1 + }}{p_{ + 1}}{p_{2 + }}{p_{ + 2}}}(A_{1} - A_{2}) \\
& = \frac{(p_{11}p_{22} - p_{12}p_{21})}{\sqrt {{p_{1 + }}{p_{ + 1}}{p_{2 + }}{p_{ + 2}}}} \\
& = \sqrt {{p_{1 + }}{p_{ + 1}}{p_{2 + }}{p_{ + 2}}} (G_{1}^2 - G_{2}^2 )
\end{split}
\end{equation*}

\section{Genesis of Canonical Correlations and Pearson $\phi$}
Given any 2x2 matrix A there exist two orthogonal matrices L and M each of order 2x2 such that
\[{{\mathop{\rm LAM}\nolimits} ^T} = \left( {\begin{array}{*{20}{c}}
{{\rho _1}}&0\\
0&{{\rho _2}}
\end{array}} \right)\]
where ${\rho _1}\left( { \ge 0} \right)$ and ${\rho _2}\left( { \ge 0} \right)$ are the singular values of the matrix A with a conventional ordering of ${\rho _1} \ge {\rho _2} \ge 0$ As a matter of fact, ${\rho _1}^2$ and ${\rho _2}^2$ are the eigenvalues of 
${{\mathop{\rm AA}\nolimits} ^T}$ and the singular values are the non-negative square root of the eigenvalues.
Let the bivariate binary distribution along with the marginals be given by
\[{\mathop{\rm Q}\nolimits}  = \left( {\begin{array}{*{20}{c}}
{{p_{11}}}&{{p_{12}}}&{{p_{1 + }}}\\
{{p_{21}}}&{{p_{22}}}&{{p_{2 + }}}\\
{{p_{ + 1}}}&{{p_{ + 2}}}&1
\end{array}} \right)\]
Let $${\mathop{\rm B}\nolimits}  = \left( {\begin{array}{*{20}{c}}
{\frac{{{p_{11}}}}{{\sqrt {{p_{1 + }}{p_{ + 1}}} }}}&{\frac{{{p_{12}}}}{{\sqrt {{p_{1 + }}{p_{ + 2}}} }}}\\
{\frac{{{p_{21}}}}{{\sqrt {{p_{2 + }}{p_{ + 1}}} }}}&{\frac{{{p_{22}}}}{{\sqrt {{p_{2 + }}{p_{ + 2}}} }}}
\end{array}} \right)$$
The singular values ${\rho _1}$ and ${\rho _2}$ of ${\mathop{\rm B}\nolimits}$ are called canonical correlations of X and Y. It turns out that ${\rho _1}=1$  and ${\rho _2}=\rho$ has the property that $1 \ge \rho  \ge 0$.
The canonical correlation $\rho$ characterizes independence of X and Y. That is 
$\rho=0$ if and only if X and Y are independent \citep{lancaster20020}. 
We do not follow this definition of canonical correlation. Technically, $\rho$ is taken to be the non-negative square root of one of the eigenvalues of ${{\mathop{\rm BB}\nolimits} ^T}$.
As a matter of fact, one of the eigenvalues is always equal to one. The other one is given by
\[{\rho ^2} = \frac{{p_{11}^2}}{{\sqrt {{p_{1 + }}{p_{ + 1}}} }} + \frac{{p_{12}^2}}{{\sqrt {{p_{1 + }}{p_{ + 2}}} }} + \frac{{p_{21}^2}}{{\sqrt {{p_{2 + }}{p_{ + 1}}} }} + \frac{{p_{22}^2}}{{\sqrt {{p_{2 + }}{p_{ + 2}}} }} - 1.\]
We want to admit both positive and negative square roots of $\rho ^2$. We have discovered that the following takes both positive and negative values in $\left[ { - 1,1} \right]$ whose square is $\rho ^2$: 
\[\rho  = \sqrt {{p_{1 + }}{p_{ + 1}}{p_{2 + }}{p_{ + 2}}} \left[ {\frac{{{p_{11}}}}{{\sqrt {{p_{1 + }}{p_{ + 1}}} }} - \frac{{{p_{12}}}}{{\sqrt {{p_{1 + }}{p_{ + 2}}} }} - \frac{{{p_{21}}}}{{\sqrt {{p_{2 + }}{p_{ + 1}}} }} + \frac{{{p_{22}}}}{{\sqrt {{p_{2 + }}{p_{ + 2}}} }}} \right].\]
We keep the same notation $\rho$. One can check that $\rho = \phi$ \citep{dunlap2000canonical}.

\section*{Results}
The power of the tests based on $Z_{1}$ and $Z_{2}$ is compareed graphically under the 100 randomly generated a bivariate distribution of X and Y (Figure 3.1). The numerical results are presented in Appendix.\\
Structurally, the graphs are similar, even though the true values of ln(odds ratio) and Person $\phi$  are on a different scale. As $ln(\theta)$ and $\rho$ moves away from the null value, the powers rise steeply towards 100 percent as expected. The spline model provides information of the underlying smoothness of power as a function of the measures of association. Similar comments do apply to the tests of Rao. A comprehensive comparison of the 4 tests is provided and in Figures 3.1 and 3.2\\  

\begin{figure}[!ht]
  \centering
  \begin{minipage}[b]{0.50\textwidth}
    \includegraphics[width=\textwidth]{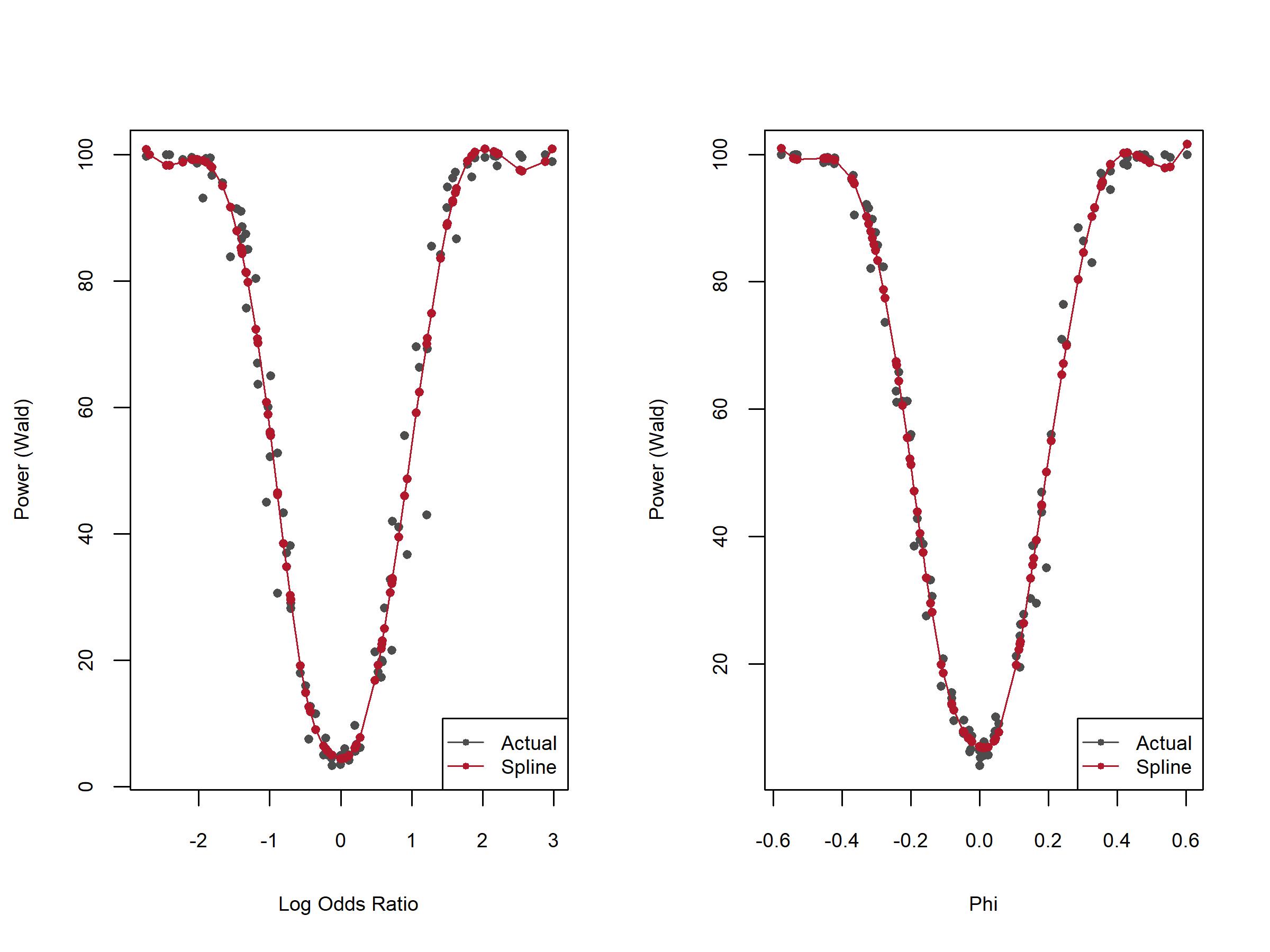}
    \caption{Wald tests: Odds ratio and canonical correlation  }
  \end{minipage}
\end{figure} 

\begin{figure}[!ht]
  \centering
  \begin{minipage}[b]{0.50\textwidth}
    \includegraphics[width=\textwidth]{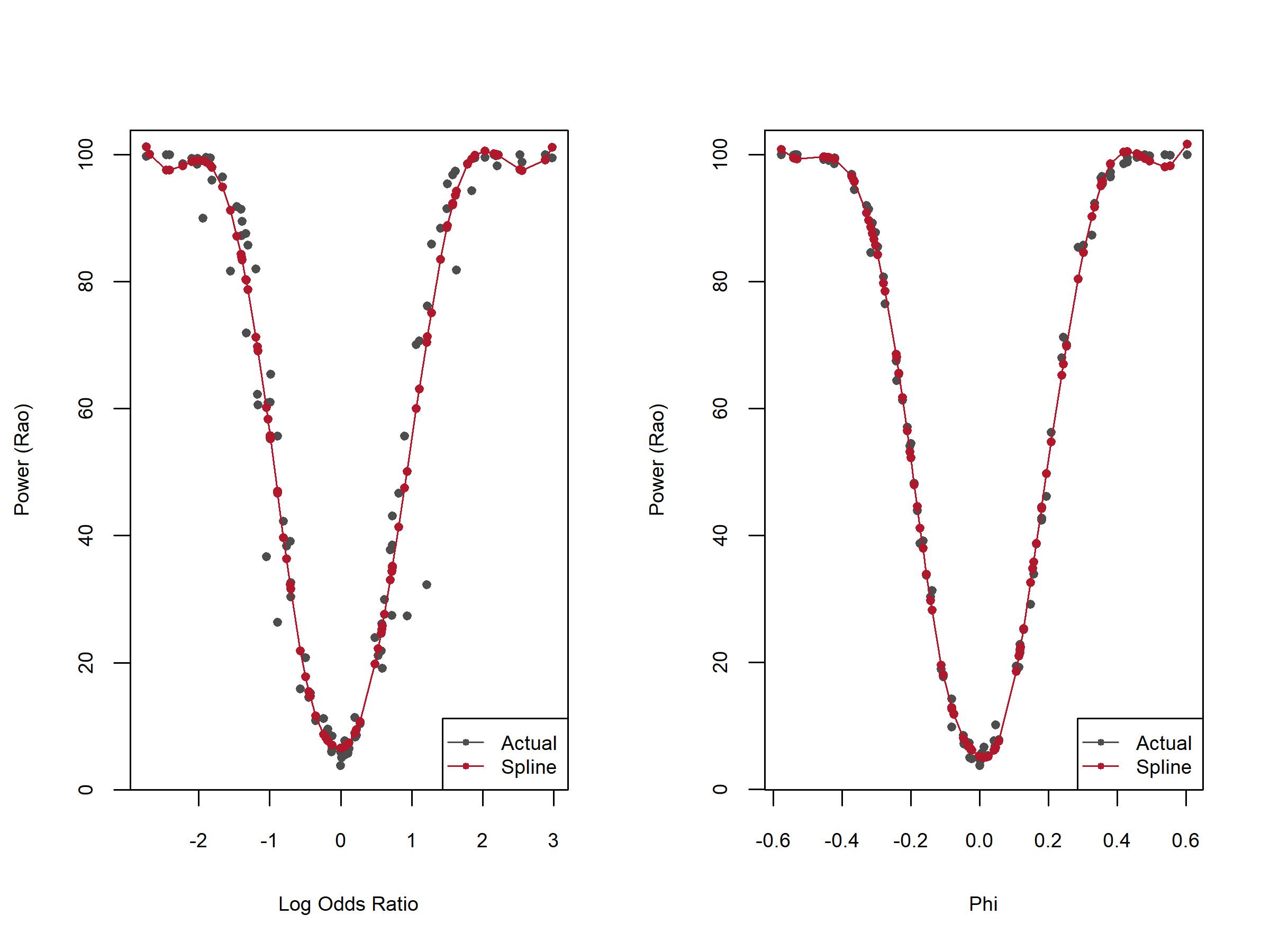}
    \caption{Rao tests: Odds ratio and canonical correlation }
  \end{minipage}
\end{figure}


Each diagonal graph is a density histogram describing the distribution of power associated with one test. Structurally, the histograms are similar meaning that the distributions are similar. Every graph below the diagonals gives the scatter plot of a pair of powers coming from two different tests with a regression line drawn on the scatter plot. Power pairs do lie more or less on the line. The graphs above the diagonal line give a Pearson correlation coefficient of the two power series. For Figure 3.3, we have generated 100 bivariate distributions of X and Y from the Uniform Dirichlet distribution on the simplex. For each distribution generated, $ln(Odds ratio)$ and Pearson $\phi$ was calculated. The scatter plot presented in Figure 3.3.\\

\begin{figure}[!ht]
  \centering
  \begin{minipage}[b]{0.50\textwidth}
    \includegraphics[width=\textwidth]{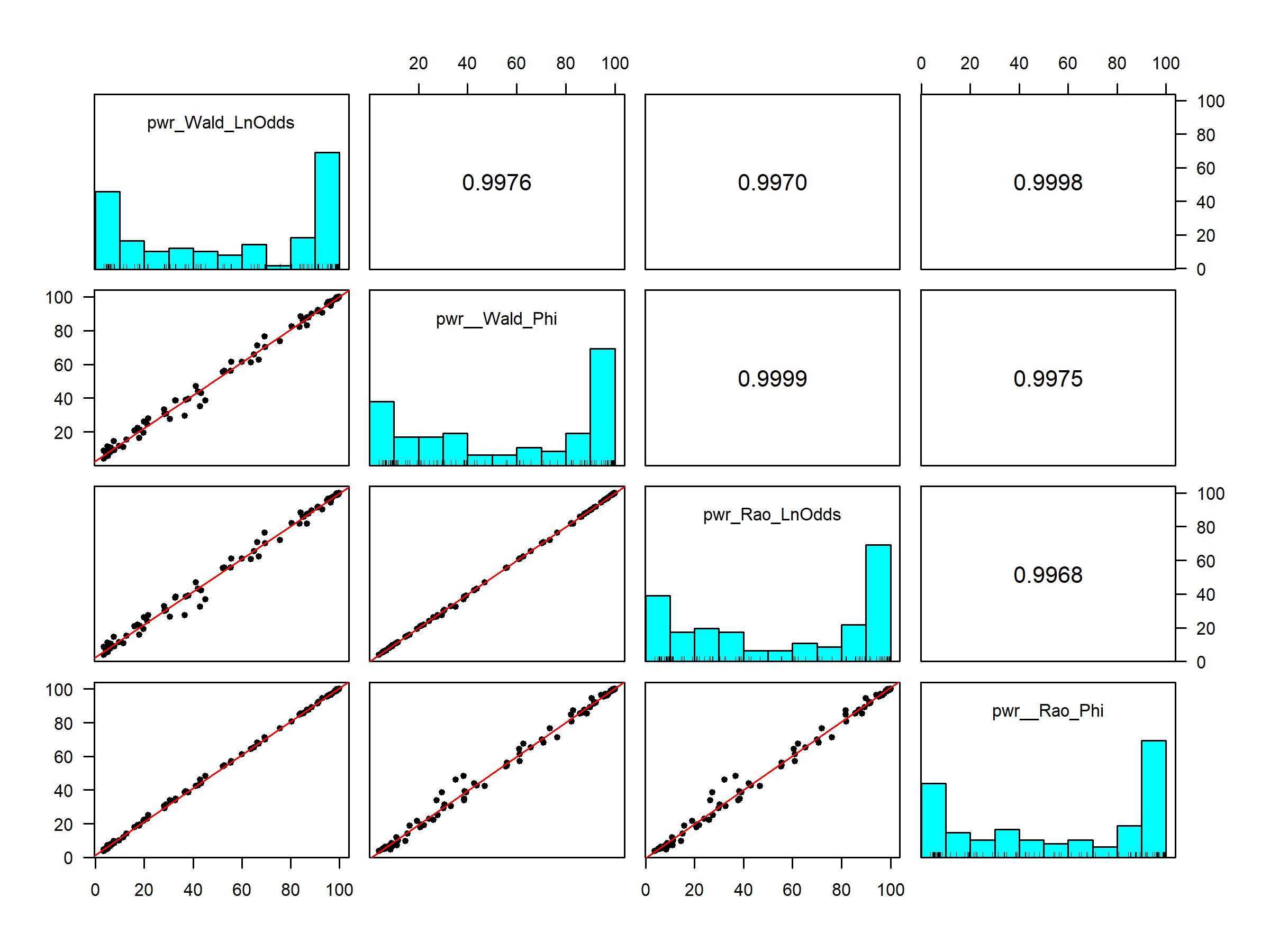}
    \caption{Correlation plots and histograms of powers }
  \end{minipage}
\end{figure}

\section{Discussion:}
For testing independence of two binary variables, we examined the power of tests built upon ln(Odds ratio) and Pearson $\phi$ (Canonical correlation $\rho$) due to Wald and Rao. These tests use asymptotic variance formulas. Our comparisons are based on a random selection of bivariate distributions from the uniform Dirichlet distribution on the simplex of bivariate distributions. We suggest that any of the four tests use in large samples.\\ 
A challenging task would be the determination of sample size for given level, power, and alternative values of the measure of association choice.
There are pros and cons in using any measure of association for testing independence. The ln(Odds ratio) has an infinite range and confidence intervals based on Odds ratio could be very wide to interpret meaningfully. Pearson $\phi$ does not have this problem. The Odds ratio does not extend beyond the  2x2 case, where   Pearson $\Phi$ is extendable to higher dimensional contingency tables. In case-control studies, the primary focus is testing equality of proportions of subjects achieving a cure. The odds ratio is used in this scenario, but Pearson $\phi$ or canonical correlation $\rho$ are inappropriate to use in such a context.\\
We have shown that Rao scores statistic based on Pearson $\phi$ is related to the traditional $\chi^{2}$ statistic of independence. Thus the $\chi^{2}$ statistic is in the ambit of the main theme of the paper.

\section*{conflict of interest}
There is no conflict of interest.


\bibliographystyle{plainnat}
\bibliography{main}

\section{Appendix 1}
\subsection{Asymptotic variance of the likelihood estimator of Pearson $\phi$}
Asymptotic variance of the  maximum likelihood estimator of Pearson correlation $\phi$
Steps:
\begin{enumerate}
\item  Joint distribution of X and Y 
$$
Q=
\begin{pmatrix}
{a} & {b}  \\
{c} & {d} \\
\end{pmatrix}
$$
\item Pearson correlation 
\begin{equation*} 
\begin{split}
\rho  & = \frac{ad - bc}{\sqrt (a+b)(a+c)(c+d)(b+d) }   \\
& = \phi \\
& = UV^{-0.5}, \mbox{ where, U = ad - bc and V = (a+b)(a+c)(c+d)(b+d) }
\end{split}
\end{equation*}
\item Generate data 
$$ 
D=
\begin{pmatrix}
n_{11} & n_{12}  \\
n_{21} & n_{22} \\
\end{pmatrix}
$$
\item  Estimator of Q ,
 $$ 
\widehat Q=
\begin{pmatrix}
\frac{n_{11}}{n} & \frac{n_{12}}{n} \\
\frac{n_{21}}{n} & \frac{n_{22}}{n}  \\
\end{pmatrix}
$$

For ease, in the description of the  asymptotic formula, use a simple notation for the entries of $\widehat{Q}$  
$$ 
\widehat Q=
\begin{pmatrix}
 j & k \\
 l & m  \\
\end{pmatrix}
$$

\item  Estimate of $\rho$,
\begin{equation*} 
\begin{split}
\widehat{\rho}  & = \frac{jm - lk}{\sqrt (j+k)(j+l)(l+m)(k+m) }   \\
& = f(j,k,l,m) \\
& = x.y^{-0.5}, \mbox{ where, x = jm - lk and V = (j+k)(j+l)(l+m)(k+m) } 
\end{split}
\end{equation*}
\item  Asymptotic variance of  $\widehat \rho $ using the delta method evaluated at  their expectations, $j= E(j), k= E(k), l= E(l), m= E(m)$
\begin{equation*}
\begin{split}
Asymptotic Variance &= \left(\frac{df}{dj} \right) ^{2} * var(j)  + 
\left(\frac{df}{dk} \right) ^{2} * var(k) + \left(\frac{df}{dl} \right) ^{2} * var(l) +  \\ &  \left(\frac{df}{dm} \right) ^{2} * var(m) + 
2 \left ( \frac{df}{dj} \right ) * \left ( \frac{df}{dk} \right) * cov(j,k)  +  \\ &
2 \left ( \frac{df}{dj} \right ) * \left ( \frac{df}{dl} \right) * cov(j,l) + 
2 \left ( \frac{df}{dj} \right ) * \left ( \frac{df}{dm} \right) * cov(j,m) + \\ &
2 \left ( \frac{df}{dk} \right ) * \left ( \frac{df}{dl} \right) * cov(k,l) + 
2 \left ( \frac{df}{dk} \right ) * \left ( \frac{df}{dm} \right) * cov(k,m) + \\ &
2 \left ( \frac{df}{dl} \right ) * \left ( \frac{df}{dm} \right) * cov(l,m)
\end{split}
\end{equation*}
 \item  Calculate the variances and covariances,  
\[\begin{array}{l}
{\mathop{\rm var}} \left( j \right) = \frac{{a\left( {1 - a} \right)}}{n};{\mathop{\rm var}} \left( k \right) = \frac{{b\left( {1 - b} \right)}}{n}\\
{\mathop{\rm var}} \left( l \right) = \frac{{c\left( {1 - c} \right)}}{n};{\mathop{\rm var}} \left( m \right) = \frac{{d\left( {1 - d} \right)}}{n}
{\mathop{\rm cov}} \left( {j,k} \right) =  - \frac{{ab}}{n};{\mathop{\rm cov}} \left( {j,l} \right) =  - \frac{{ac}}{n}\\
{\mathop{\rm cov}} \left( {j,m} \right) =  - \frac{{ad}}{n};{\mathop{\rm cov}} \left( {k,l} \right) =  - \frac{{bc}}{n}\\
{\mathop{\rm cov}} \left( {k,m} \right) =  - \frac{{bd}}{n};{\mathop{\rm cov}} \left( {l,m} \right) =  - \frac{{cd}}{n}
\end{array}\]
\item 
\begin{equation*} 
\begin{split}
\frac{df}{dj}  & = x \left ( \frac{dy^{-0.5}}{dj} \right )  + y^{- 0.5}  \left ( \frac{dx}{dj} \right)  \\
& = x(-0.5)y^{- \frac{3}{2}} \frac{dy}{dj} + y ^{- 0.5} \left ( \frac{dx}{dj} \right)  \\
& = -(0.5)xy ^{- 0.5}y ^{- 1}(2j+k+l)(l+m)(k+m) + y ^{- 1}m    
\end{split}
\end{equation*}
\item 
\[\begin{array}{c}
{\left( {\frac{{\partial f}}{{\partial j}}} \right)_{j = {\mathop{\rm E}\nolimits} \left( j \right),k = {\mathop{\rm E}\nolimits} \left( k \right),l = {\mathop{\rm E}\nolimits} \left( l \right),m = {\mathop{\rm E}\nolimits} \left( m \right)}} =  - {\textstyle{1 \over 2}}u{v^{ - {\raise0.5ex\hbox{$\scriptstyle 1$}
\kern-0.1em/\kern-0.15em
\lower0.25ex\hbox{$\scriptstyle 2$}}}}{v^{ - 1}}\left( {2a + b + c} \right)\left( {c + d} \right)\left( {b + d} \right) + {v^{ - {\raise0.5ex\hbox{$\scriptstyle 1$}
\kern-0.1em/\kern-0.15em
\lower0.25ex\hbox{$\scriptstyle 2$}}}}d\\
 =  - {\textstyle{1 \over 2}}\rho {v^{ - 1}}\left( {2a + b + c} \right)\left( {c + d} \right)\left( {b + d} \right) + {v^{ - {\raise0.5ex\hbox{$\scriptstyle 1$}
\kern-0.1em/\kern-0.15em
\lower0.25ex\hbox{$\scriptstyle 2$}}}}d
\end{array}\]
\item 
\[{\left( {\frac{{\partial f}}{{\partial k}}} \right)_{j = {\mathop{\rm E}\nolimits} \left( j \right),k = {\mathop{\rm E}\nolimits} \left( k \right),l = {\mathop{\rm E}\nolimits} \left( l \right),m = {\mathop{\rm E}\nolimits} \left( m \right)}} =  - {\textstyle{1 \over 2}}\rho {v^{ - 1}}\left( {2b + a + d} \right)\left( {a + c} \right)\left( {c + d} \right) - {v^{ - {\raise0.5ex\hbox{$\scriptstyle 1$}
\kern-0.1em/\kern-0.15em
\lower0.25ex\hbox{$\scriptstyle 2$}}}}c\]
\item
\[{\left( {\frac{{\partial f}}{{\partial l}}} \right)_{j = {\mathop{\rm E}\nolimits} \left( j \right),k = {\mathop{\rm E}\nolimits} \left( k \right),l = {\mathop{\rm E}\nolimits} \left( l \right),m = {\mathop{\rm E}\nolimits} \left( m \right)}} =  - {\textstyle{1 \over 2}}\rho {v^{ - 1}}\left( {2c + a + d} \right)\left( {a + b} \right)\left( {b + d} \right) - {v^{ - {\raise0.5ex\hbox{$\scriptstyle 1$}
\kern-0.1em/\kern-0.15em
\lower0.25ex\hbox{$\scriptstyle 2$}}}}b\]
\item
\[{\left( {\frac{{\partial f}}{{\partial m}}} \right)_{j = {\mathop{\rm E}\nolimits} \left( j \right),k = {\mathop{\rm E}\nolimits} \left( k \right),l = {\mathop{\rm E}\nolimits} \left( l \right),m = {\mathop{\rm E}\nolimits} \left( m \right)}} =  - {\textstyle{1 \over 2}}\rho {v^{ - 1}}\left( {2b + b + c} \right)\left( {a + b} \right)\left( {a + c} \right) + {v^{ - {\raise0.5ex\hbox{$\scriptstyle 1$}
\kern-0.1em/\kern-0.15em
\lower0.25ex\hbox{$\scriptstyle 2$}}}}a\]
\item
The expression derived in steps 1 through 12 are plugged into the asymptotic variance formula in Step 6.
\item  if $\rho = 0$ then Asymptotic variance $\left(\widehat{\rho} \right) = \frac{1}{n}$
\end{enumerate}

\end{document}